\documentclass[twocolumn]{aastex63}

\usepackage{enumitem}

\shorttitle{Floofs}
\shortauthors{Mayorga et al.}
\graphicspath{{./}{figures/}}

\begin{document}

\title{Detection of Rotational Variability in Floofy Objects at Optical Wavelengths}

\correspondingauthor{L. C. Mayorga}
\email{laura.mayorga@jhuapl.edu}

\author[0000-0002-4321-4581]{L.~C. Mayorga}
\author[0000-0002-2739-1465]{E. M. May}
\author{J. Lustig-Yaeger}
\affiliation{The Johns Hopkins University Applied Physics Laboratory 11100 Johns Hopkins Rd Laurel, MD, 20723, USA}

\author{S. E. Moran}
\affiliation{Department of Earth and Planetary Sciences, Johns Hopkins University, 3400 N. Charles St, Baltimore, MD 21218, USA}

\begin{abstract}
Phase resolved observations of planetary bodies allow us to understand the longitudinal and latitudinal variations that make each one unique. Rotational variations have been detected in several types of astronomical bodies beyond those of planetary mass, including asteroids, brown dwarfs, and stars. Unexpected rotational variations, such as those presented in this work, reminds us that the universe can be compli\textit{cat}ed, with more mysteries to uncover. In this work we present evidence for a new class of astronomical objects we identify as ``floofy" with observational distinctions between several sub-types of these poorly understood objects. Using optical observations contributed by the community, we have identified rotational variation in several of these floofy objects, which suggests that they may have strong differences between their hemispheres, likely caused by differing reflectivity off their surfaces. Additional sub-types show no rotational variability suggesting a uniform distribution of reflective elements on the floofy object. While the work here is a promising step towards the \textit{cat}egorization of floofy objects, further observations with more strictly defined limits on background light, illumination angles, and companion objects are necessary to develop a better understanding of the many remaining mysteries of these astronomical objects.
\end{abstract}


\section{Introduction} \label{sec:intro}
It is well known in the solar system that the viewing geometry of an observation determines the amount of reflected light detected from an object. This has typically been modeled using a bidirectional reflectance function, such as that described in \citet{Hapke2012}, or a single particle phase function such as the one-term Henyey-Greenstein phase function, 
\begin{equation}
 p(g) = \frac{1-b^2}{(1+2b\cos{g}+b^2)^{2/3}}  
\end{equation}
where $b$ is the asymmetry factor and varies between -1 and 1, where negative values represent a backward s\textit{cat}tering surface and positive values forward s\textit{cat}tering \citet{Henyey1941}. These are dependent on the phase angle, the emission angle, and the incident angle on a particular surface. 

The measurement of optical light curves has been particularly well used in the study of asteroids and the other small bodies in our solar system, e.g., irregular satellites and asteroids. For small bodies, these variations are generally due to elongated shape, more akin to the ellipsoidal variations seen in stars under extreme tidal forcing from a nearby massive object. The brightness range is thus related to how elongated the body is and how the periodicity relates to the rotation period. The best-fit shapes, combined with the knowledge of the spin period of the objects, can allow one to estimate the bulk densities of these objects \citep{Lacerda2007}. However, this is only true when the body is fairly uniform in brightness and composition.

Planets, however, are frequently not uniform. From Jupiter observations, we already have reason to suspect that the typical image of Jupiter's red and white bands of clouds does not hold true when Jupiter is viewed from its poles. Indeed, even terrestrial planets such as Earth and Mars have polar ice caps where the compositional ices have different s\textit{cat}tering properties and would affect polar observations of either planet. Even the dwarf planet Pluto shows significant albedo variance from equator to poles \citep{buratti2017}, due to compositional differences across the surface \citep{grundy2016}.

While Solar System observations allow us to study planets from a variety of viewing geometries through direct missions to these worlds, exoplanets provide a unique, and perhaps infuriating, case study in which we do not always know the angles at which we observe a planet. Hot Jupiters on circular (non-eccentric) orbits are generally assumed to be tidally locked to their host stars since the timescales over which synchronous rotation occurs are shorter than timescales over which an orbit circularizes \citep{Rasio1996}. Tidally locked planets should further have a negligible axial obliquity, implying that in these cases, we can assume the latitude at which we observe the planet based on the inclination of the planet's orbit. However for those on longer orbits, it is entirely possible that they are not tidally locked, and therefore may have a significant obliquity to contend with, suggesting we cannot assume the latitude at which we are observing the planet based simply on the viewing geometry of the plane of the planet's orbit. For these non-tidally locked planets, it is extremely difficult to constrain the rotation rate, leading to degeneracies when hoping to determine which side of a planet you are observing during full orbit observations. 

Direct imaging observations of faint objects with next-generation space-based observatories (e.g., CatEx) will face noise dominated by leakage through a coronagraph given as
\begin{equation}
 c_{speck} \propto C A \mathcal{T}  
\end{equation}
where $C$ is the design contrast of the coronagraph, $A$ is the light collecting area, and $\mathcal{T}$ is the end-to-end throughput of the opti\textit{cat} system. Understanding these noise sources will allow us to study the nuances in the light curves of small bodies and planetary objects alike.

Beyond the detection of hot Jupiters with direct imaging, background source contamination and miss-classifi\textit{cat}ion of observed objects are pressing issues. One suggestion is to use color rather than just brightness, and a variety of filter combinations have been tested and evaluated for differentiating planets \citep{Cahoy2010,Hegde2013,Mayorga2016, Batalha2018}. Further work on has shown that it may be possible to differentiate between low and high density worlds using differences between eclipses and transits in specific colors \citep{May2020b,Sotzen2021}. Typically, in gas giant planets the blue optical slope is a product of Rayleigh s\textit{cat}tering, while red optical and near-infrared spectral features are due to a variety of gas species. The presence of high altitude condensates or photochemical hazes introduces significant degeneracies between low and high density objects, though for significantly smaller objects this effect does not appear to be an issue in existing observations \citep{moran2018}.  Thus the appropriate selection color and observing filter is necessary at the outset of observations. 

Recent observational evidence at optical wavelengths has presented a new class of potentially rotationally variable objects with tentative variable reflectivity as a function of viewing angle, similar to the variance seen in Solar System planets and exoplanets. While it remains to be seen where this new class of \textit{cat}stronomical bodies -- which we henceforth refer to as ``floofy objects" (sometimes referred to as UFOs: understudied floofy objects) -- falls in the story of planets, brown dwarfs, asteroids, stars, and other rotationally variable objects, we are confident that they will provide an unending amount of joy as astronomers work to understand their unique peculiarities.

In this paper, we present an exploratory study into the phase variations detected in floofy objects at optical wavelengths with a particular focus on the observed belly to non-belly transition. In \autoref{sec:data}, we present the data collection of our floofy objects. In \autoref{sec:and} we discuss our data analysis methods and present the results in \autoref{sec:res}. We discuss the impli\textit{cat}ions of our results on the study of small bodies and exoplanets in \autoref{sec:disc}, and present a summary and our conclusions in \autoref{sec:concl}.

\section{Data Collection and Selection} \label{sec:data}
\begin{table*}[]
    \centering
    \footnotesize
    \begin{tabular}{|c|c|c|c||c||c|c|c|c|c|c|c|c|}
    \hline 
         \textbf{Sub-Type} & \textbf{Description} & \textbf{\# Observed} & \textbf{\% Used} & \textbf{\# Frames} & \textbf{$\pm$ 180} & \textbf{-135} & \textbf{-90} & \textbf{-45} & \textbf{0} & \textbf{45} & \textbf{ 90} & \textbf{135} \\ \hline
         BK & Black & 25 & 28.0 & 17 & 0 & 2 & 2 & 1 & 6 & 3 & 2 & 1 \\
        BW & Black \& White & 54 & 48.15 & 56 & 4 & 5 & 6 & 11 & 14 & 5 & 5 & 6 \\
        GR* & Grey & 9 & 11.11 & 5 & 0 & 1 & 1 & 0 & 1 & 1 & 1 & 0 \\
        GW* & Grey \& White & 18 & 50.0 & 24 & 5 & 1 & 1 & 1 & 8 & 4 & 3 & 1 \\
        OG & Orange & 19 & 42.11 & 14 & 2 & 1 & 2 & 1 & 5 & 0 & 2 & 1 \\
        OW & Orange \& White & 24 & 41.67 & 21 & 1 & 2 & 2 & 2 & 8 & 0 & 4 & 2 \\
        WT & White & 12 & 50.0 & 18 & 2 & 1 & 2 & 2 & 5 & 3 & 2 & 1 \\
        TB & Tabby & 105 & 45.71 & 66 & 7 & 3 & 9 & 7 & 21 & 6 & 6 & 7 \\
        TW* & Tabby \& White & -- & -- & 43 & 5 & 2 & 6 & 1 & 14 & 7 & 6 & 2 \\
        CL & Calico & 40 & 57.5 & 61 & 8 & 4 & 5 & 6 & 17 & 9 & 6 & 6 \\
        TS & Tortie & 14 & 57.14 & 17 & 3 & 0 & 2 & 3 & 6 & 2 & 1 & 0 \\
        RD & Rag Doll & 6 & 16.67 & 2 & 1 & 0 & 0 & 0 & 1 & 0 & 0 & 0 \\
        SM & Siamese & 1 & 100.0 & 2 & 0 & 0 & 1 & 0 & 1 & 0 & 0 & 0 \\ \hline
         { } & TOTALS & 328 & 45.12 & 346 & 38 & 22 & 39 & 35 & 107 & 40 & 38 & 27  \\ \hline
    \end{tabular}
    \caption{Overview of observed floofy objects \textit{cat}egorized by defined sub-types. We include both the total number of floofy objects observed and the percentage included in this study. The total number of frames per sub-type is denoted, including a breakdown by sub-observer longitude. We note that TW and GW were later additions, with their original data classifi\textit{cat}ions falling within TB and BW, respectively. The GW and GR classifi\textit{cat}ions were added halfway through the observational phase, so some number of the unused GW and GR objects may be marked as BW or BK, respectively. TW was added after data observations, so unused TW are classified as TB. For all sub-types with a significant number of observations, about half of them were deemed acceptable for the study, so we can assume this holds true for the re-classified objects as well.}
    \label{table:data}
\end{table*}

Data was independently collected by citizen \textit{cat}stronomers over a variety of observing epochs and requested via social media. We solicited for images specifying the following criteria:
\begin{enumerate}[leftmargin=0.2in]
    \setlength\itemsep{-0.7em}
    \item It was an observation of a floofy object rolling around,
    \item that we had particular interest in floofy objects with different colored belly and non-belly regions, 
    \item did they want their observations of said floofy objects used for important scientific research?
\end{enumerate}
We also provided follow up information on the request:
\begin{itemize}[leftmargin=0.2in]
    \setlength\itemsep{-0.7em}
    \item that we preferred well-lit observations with significant signal to noise (SNR) and minimal contamination from shadows or background non-floofy objects,
    \item that multiple pictures from different angles would be better in order to help understand the belly to non-belly transition in floofy objects, and
    \item that we accepted observations of transiting and eclipsing floofy objects\footnote{Original plans for this project included a study of the effects of binary objects, including searching for the rarer and illusive circumbinary floofy object. Unfortunately, the overwhelming response to the initial request impacted our ability to perform this search, though \cite{May2016} shows that there is unlikely to be a difference in the observed colors of circumbinary floofy objects and their single-star equivalents.}.
\end{itemize}

The majority of images were provided through Twitter replies to the solicitation, however some were also obtained through private communi\textit{cat}ion on other platforms, i.e. Slack. We excluded observations that were sent in the form of unsolicited direct messages to ensure the safety of participants and researchers. After an overwhelming response of the community to contributing to research on these poorly understood floofy objects, we waited at least 24 hours before beginning to filter and classify contributed data to allow the community to submit their observations. Individual frames were then filtered to exclude those that did not meet our initial criteria (i.e., dark with low SNR and/or shadows; angles that did not sufficiently show the belly to non-belly transition). The overwhelming response also resulted in an exclusion of many data sets with only a single observation of a particular floofy object. In a few cases we solicited additional frames of a particular floofy object if it was a sub-type with insufficient observations. We further excluded frames that had a significant portion of the floofy object outside of the field of view. 

Previous research on colors of floofy objects measured variations due to their underlying genetic structure that control, among other things, striping, rotational variability, and patchy colors such as spotted regions \citep[e.g.,][]{CatGenes1,CatGenes2}. Because the observations included in this work do not allow us to probe this underlying structure of floofy objects, lassifi\textit{cat}ion of floofy objects was done visually and attempted to follow these previously identified trends in colors. Guesses of sub-type were required when an object's coloring was ambiguous or unclear. 

\begin{figure}[t]
    \plotone{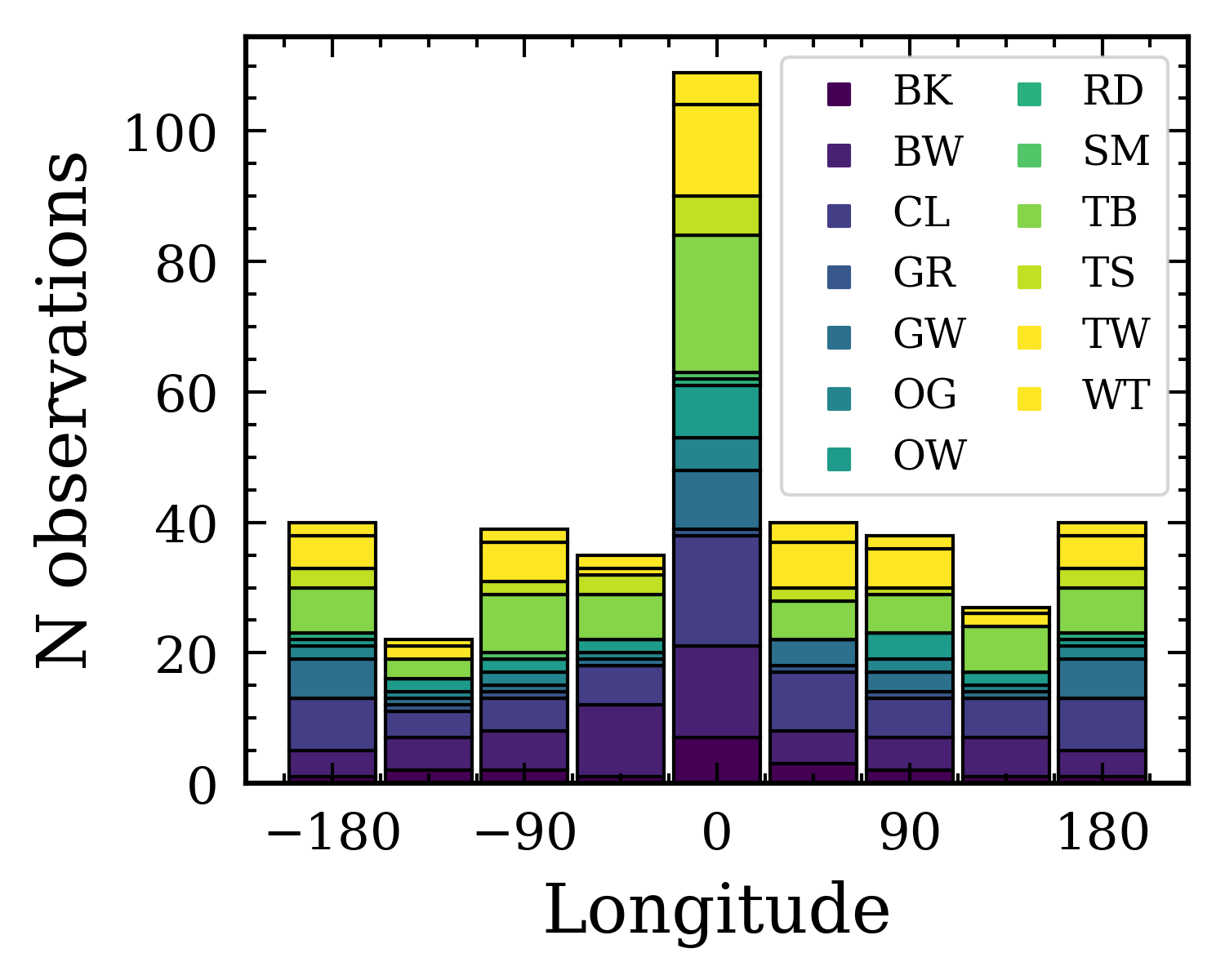}
    \caption{A histogram detailing the number of observations per sub\textit{cat}egory of object observed and longitude of observation.}
    \label{fig:obs_hist}
\end{figure}

\autoref{table:data} outlines the specifics of the data collected, including the total number of floofy objects observed and how many fell in each floofy object sub-type.  A histogram version of the number of floofy objects observed in each sub-type vs. longitude is also shown in \autoref{fig:obs_hist}. The majority of the objects were from 0$^\circ{}$ longitude. We also note the total percentage of floofy objects for which observations met our inclusion criteria in each sub-type. Sub-types were pre-defined based on the previously observed trends in these objects as noted above, and as addressed in the table caption; additional \textit{cat}egories were added as necessary. In particular, while the TB class objects are the largest in our sample, our initial classifi\textit{cat}ion did not differentiate class TB floofy objects with clear rotational variation from those without. We later determined this was a necessary differentiation (denoted as TW) but did not re\textit{cat}egorize discarded observations of the TB class, leaving us with incomplete data on the percentage of the observed TB class with rotational variability.

For the sake of brevity, we do not list the observing dates and times, but these can be requested as necessary from the individual data collectors\footnote{Raw data is available to download at \url{bit.ly/3u8dg8p}}. While exact camera specifi\textit{cat}ions are not known for most observations, we assume the majority of data collection is done with a camera on an average smart phone. Most citizen \textit{cat}stronomers who provided data are from the United States, where iOS is the dominant operating system\footnote{\url{bit.ly/3w6ysx4}}, however we are unsure if this statistic holds true for the sub-population who contributed to this data collection effort. Luckily, \cite{Burggraaff:19} find that there is visually a minimal difference in spectral response between the iPhone SE and Galaxy S8, suggesting that camera differences is likely to be a minimal source of astrophysical s\textit{cat}ter compared to the more dominant uncertainty caused by differing lighting and angles in the contributed data. Occasionally floofy objects are seen to be interacting with colorful smaller nearby companions, which we assume to be for their entertainment. While we attempt to remove this contamination, occasionally they may impact the average color of the frame. In \autoref{sec:and} we discuss our methods to color-correct frames to account for the differing seeing conditions across the observations due to background and s\textit{cat}tered light, as well as to remove nearby colorful companions.

\section{Data Analysis} \label{sec:and}
While \cite{Burggraaff:19} do show that there are flat fielding and gain differences between the two consumer cellular cameras studied, the center of the frame where most collected data appeared is shown to have relatively constant flat and gain components, therefore we do not perform any flat field correction to the data.

\begin{figure}[h!]
    \centering
    \includegraphics[width=0.95\linewidth]{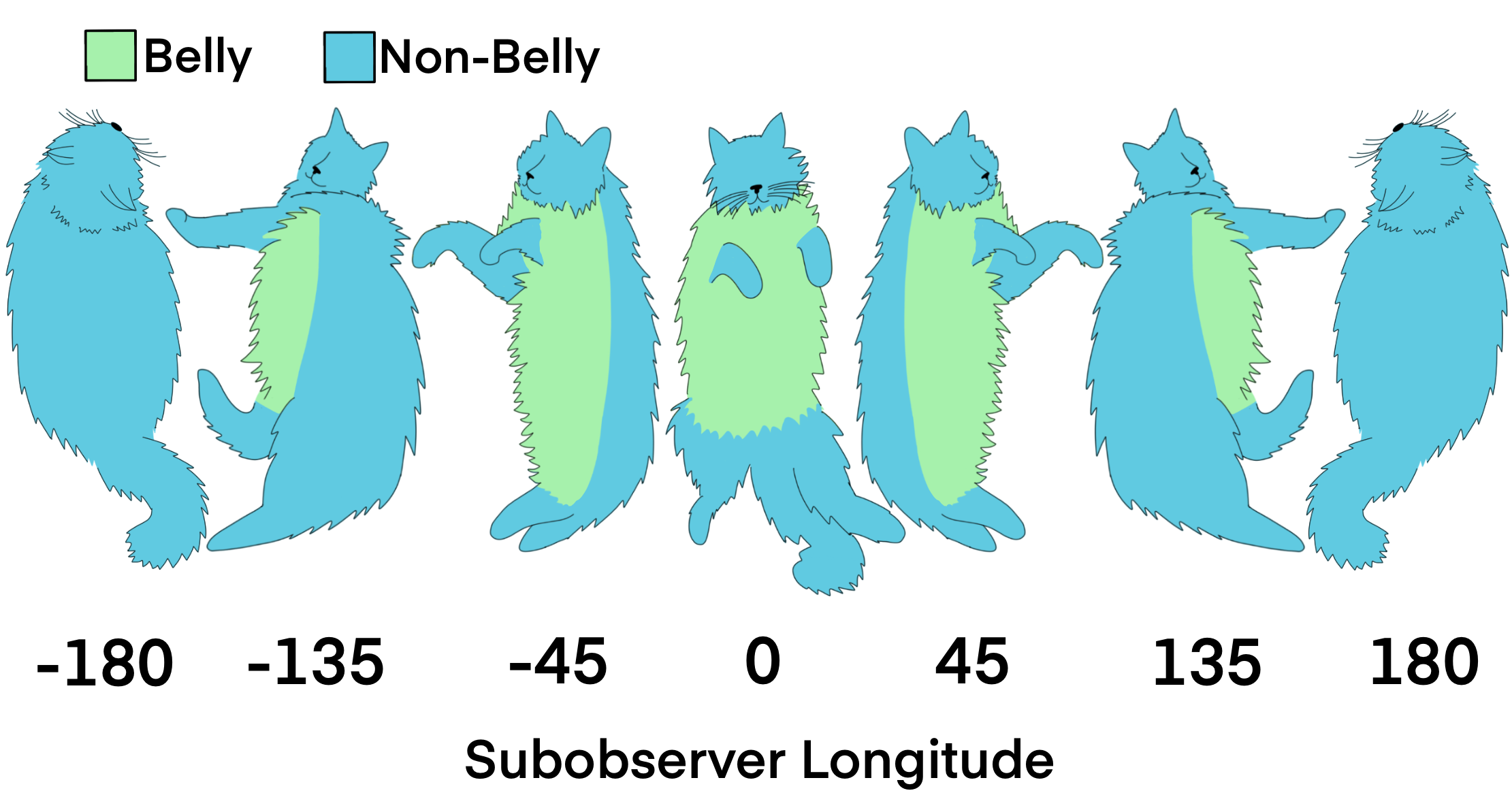}
    \includegraphics[width=0.95\linewidth]{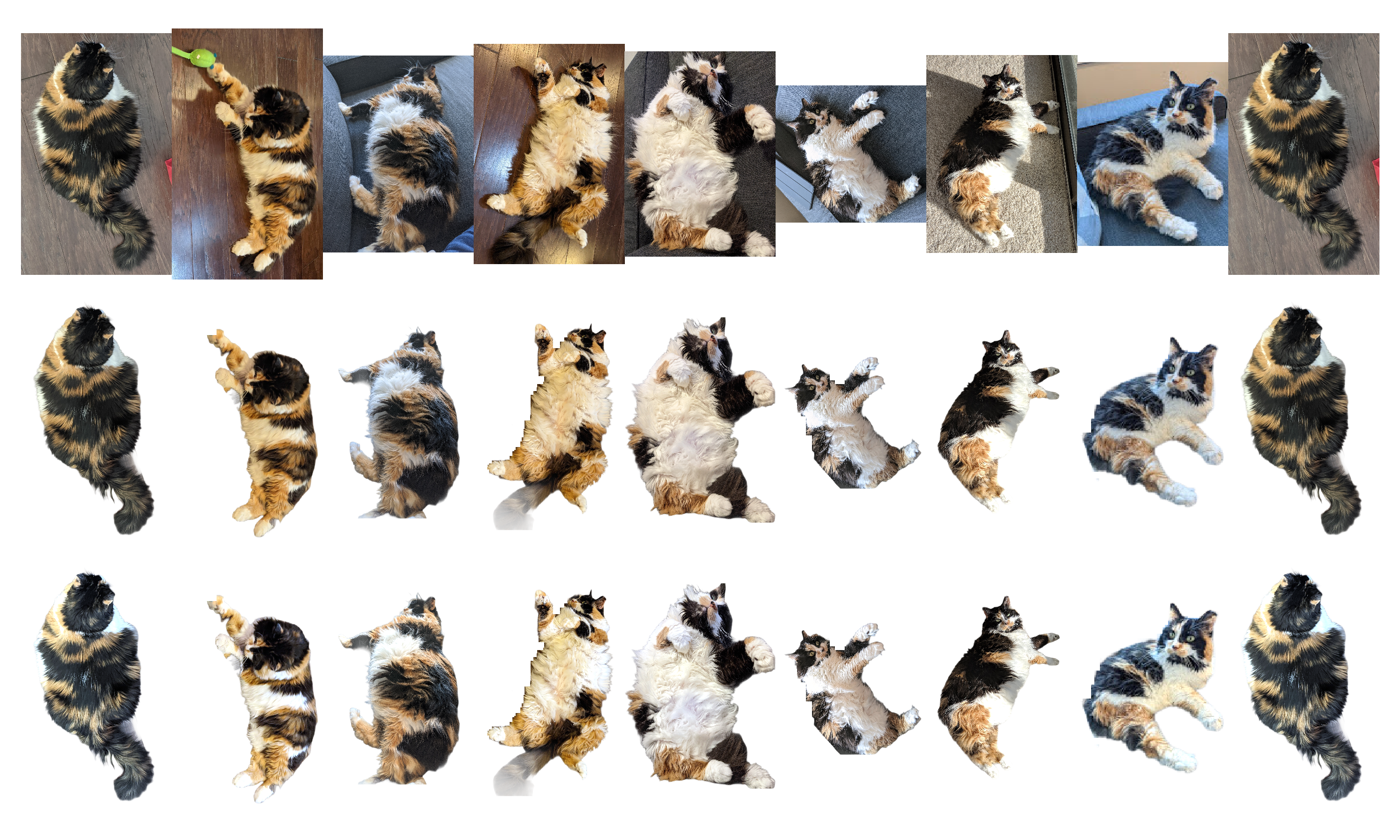}
    \caption{The longitudinal coordinate system of floofy objects and examples of image reduction. Top (first row): here we define the longitudinal coordinate system of floofy objects in units of degrees. Green denotes belly-regions and blue denotes non-belly regions. While not indi\textit{cat}ed on this figure, we further define the head of the floofy object as a sub-observer latitude of 90$^{\circ}$ and the tail of the object as a sub-observer latitude of -90$^{\circ}$. Second row: the original images of object CL1 at various longitudes. Third row: the background subtracted images of object CL1 using the automatic background subtraction routine. Bottom (4th row): The white balanced images of object CL1. Note how this has significantly balanced images near -45$^\circ{}$ longitude but has also added a blue tint to images near -180 and 180$^\circ{}$ longitude.}
    \label{fig:cat_coordinates}
\end{figure}

Background removal is done using the \url{Slazzer.com} API. Because each object is highly resolved and not rotationally symmetric, we do not compare aperture sizes and use the default background removed frames. All frames are able to successfully identify a foreground object, though in some cases background regions that appear soft and colorful (presumably for the floofy object's comfort) remain. These frames are flagged and manual removal of contaminated regions is done. Frames are then visually \textit{cat}egorized by sub-type (see \autoref{table:data}) and approximate sub-observer latitude and longitudes based on the identified radial profile of the subject. \autoref{fig:cat_coordinates} demonstrates our defined longitude grid on floofy objects. 

The assignment of sub-observer longitude is performed as follows: First, we determine the direction the floofy object is facing. If it is facing left, we assign negative sub-observer longitudes, and if it is facing right we assign positive sub-observer longitudes. Second, we visually assess the amount of belly seen on the central region (excluding all heads, tails, arms, and legs) of the floofy object. If the entire central region is determined to be belly, we assign a sub-observer longitude of 0$^{\circ}$. More than half of this region being belly is given a value of 45$^{\circ}$, approximately half of the region being belly is assigned 90$^{\circ}$, while more than half of the region being belly is assigned 135$^{\circ}$ (with the corresponding + or - sign). If no belly is present, this is denoted as a sub-observer longitude of 180$^{\circ}$.

While not shown in \autoref{fig:cat_coordinates}, we define the head to be a sub-observer latitude of 90$^{\circ}$ and the tail to be the sub-observer latitude of -90$^{\circ}$. Due to the non-spherical nature of floofy objects, for most frames we assume a sub-observer latitude of 0$^{\circ}$ such that the floofy object is being observed along the equator with equal parts of the northern and southern hemispheres visible. In rare cases we indi\textit{cat}e that the floofy object was observed from either the northern or southern pole based on the angle of the provided frames. Because we are primarily interested in the belly to non-belly transition in these objects, the additional latitudinal dimension generally does not provide useful data for understanding this transition. Further, we find that these floofy objects are less symmetric in the latitudinal dimension and therefore studying any transitions in that dimension is mostly beyond the scope of this work.

We first white balance all images to account for variations in lighting across images of the same object. We normalize the image using specific percentile values in each channel and assume that the whitest patch is set at 98\% percentile of the image intensity values. We chose this percentile for an aggressive whitening, without having too strong of a blue tint in images where it was not necessary. Following \citet{Mayorga2020}, we use a common color theory practice for image file types to determine the average brightness in each channel for the images, $\overline{B_j}$
\begin{equation}
\overline{X} = \sqrt{\sum_i \frac{x_{i}^2}{N}}
\end{equation}
where $x_{i}^2$ is the value of the $i$th pixel in the image for channel $X$, which is either $R$, $G$, or $B$, \citep[see][]{McREYNOLDS2005}.

As discussed in \cite{May2020c}, the choice of data reduction method can have a large impact on the resulting rotation curves, and a uniform appli\textit{cat}ion of data reduction method is key to determining trends in an underlying population. To those ends, while we note that individual frames may have benefited from a more specialized reduction method, we make a conscious choice to apply the same uniform correction method to each frame observed. We make our data correction and analysis publicly available\footnote{\url{https://github.com/lcmayor/floofy-fluxuations}}.

\section{Results} \label{sec:res}
\begin{figure}[!t]
    \plotone{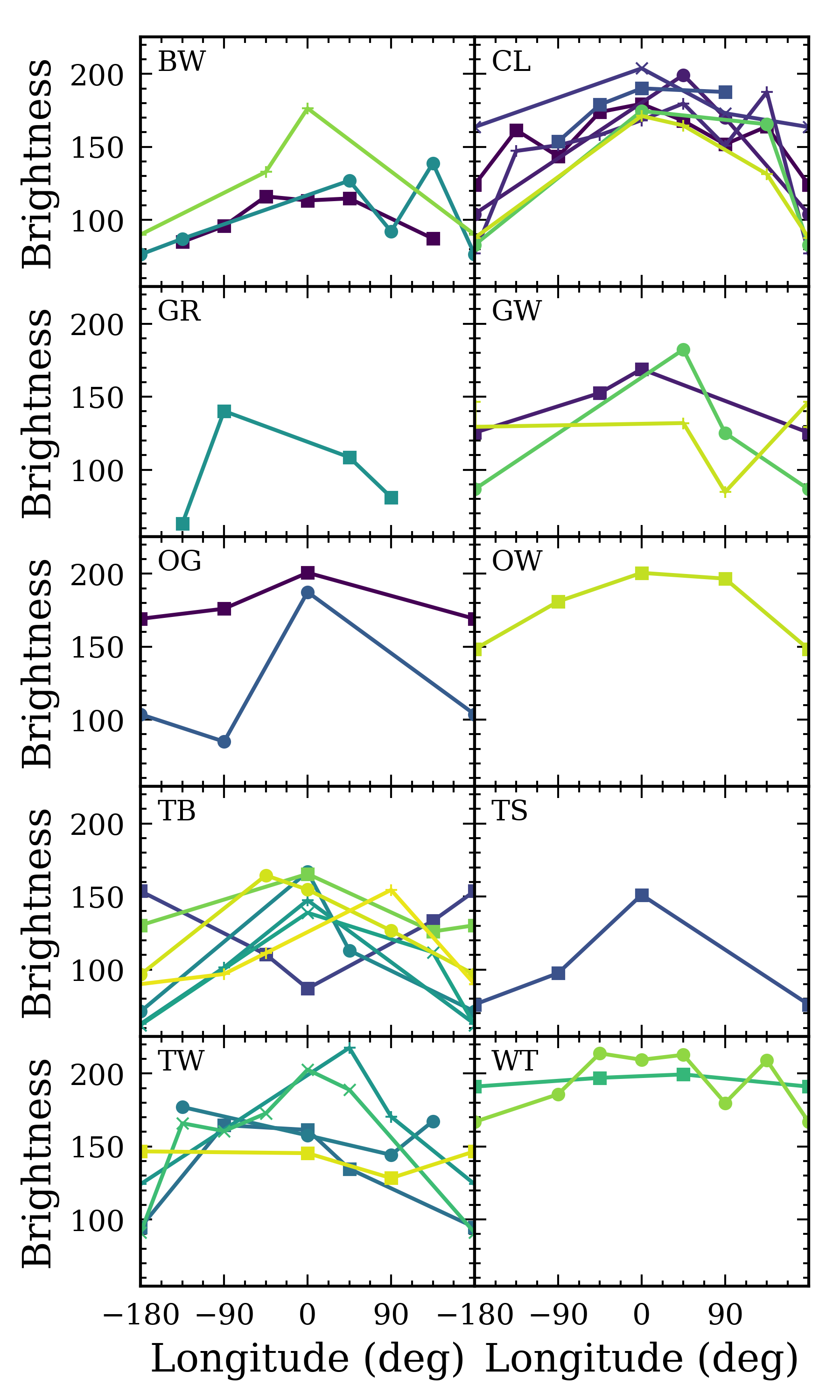}
    \caption{White light rotation curves of each object where four or more longitudes were observed. Each panel is the rotation curve for each object sub\textit{cat}egory, where each color and symbol corresponds to a different object.}
    \label{fig:white}
\end{figure}

We compute a human-perceived brightness based on the ITU-R Recommendation BT.601 \citep{international2007studio}, $Y_\mathrm{601}$,
\begin{equation}
Y_{\mathrm{601}}=0.299R+0.587G+0.114B,
\end{equation}
which we will refer to as brightness. We show the white light rotation curves for each object that had four or more longitudes observed in \autoref{fig:white}. Since for these objects -180$^\circ{}$ is equivalent to 180$^\circ{}$, we thus repli\textit{cat}e those values at -180$^\circ{}$ and 180$^\circ{}$.

We can see that typically these objects tend to be whiter and brighter at longitudes of 0 degrees, with some exceptions. In particular, sub\textit{cat}egory WT shows very little variation with longitude. Sub\textit{cat}egory BW suggests a brighter whiter longitude 0 degree point but is much weaker than other sub\textit{cat}egories, such as the more rotationally variable TW and CL.

\begin{figure}[!t]
    \plotone{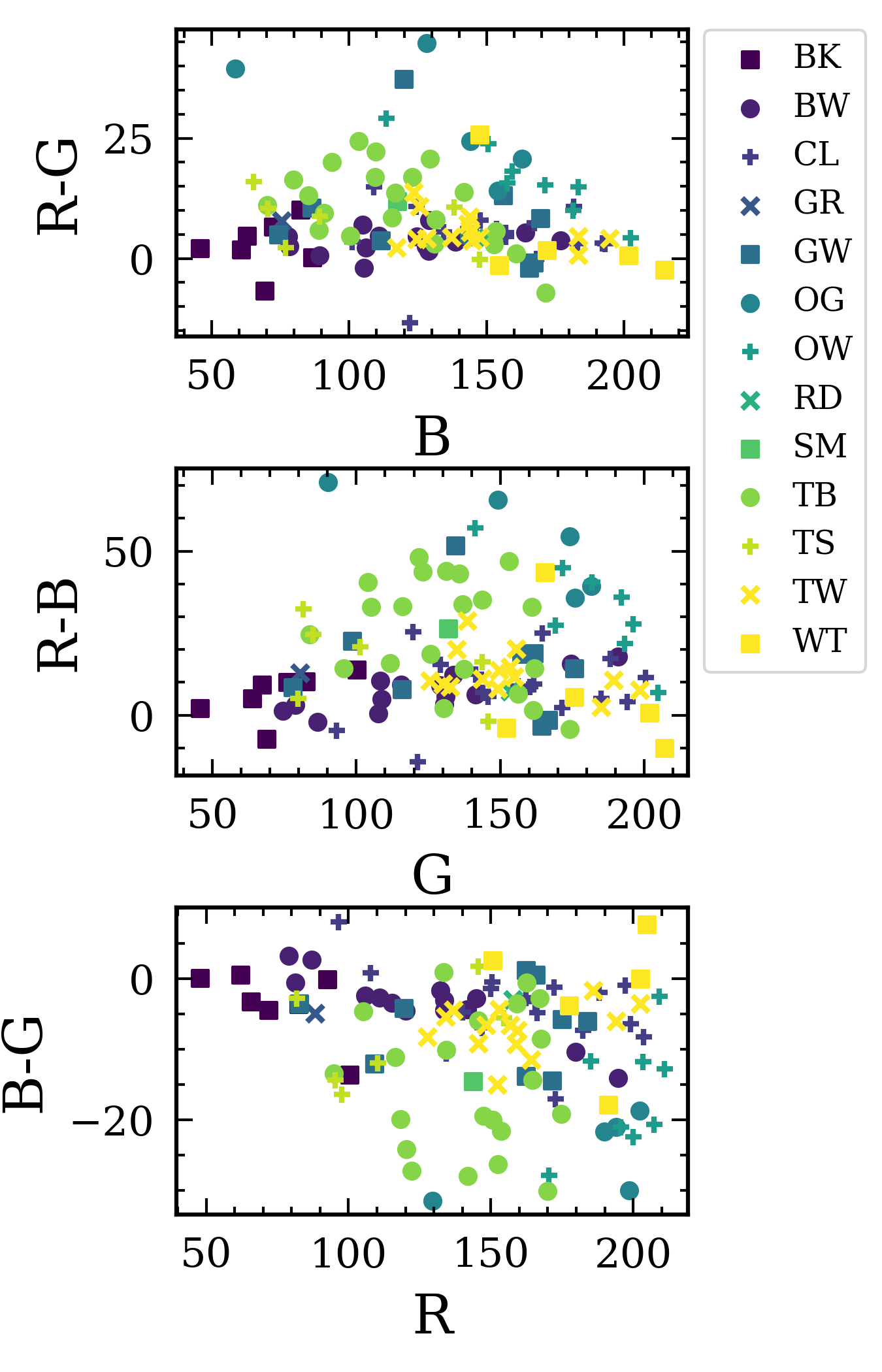}
    \caption{Color-magnitude diagrams for the three filters in which observations were made at 0$^\circ{}$ longitude. Each \textit{cat}egory is a different color and symbol.}
    \label{fig:colo_mag}
\end{figure}

We show the color-magnitude diagram including all sub\textit{cat}egories only for observations at 0$^\circ{}$ longitude in \autoref{fig:colo_mag}. As mentioned previously, sub-types were pre-defined with additional \textit{cat}egories added as necessary. We can clearly see a cluster of BK and BW trending into TW and WT across 0 in the $(X-Y)$, where X and Y can be any one of ${R, G, B}$. $(R-G)$ is most symmetric in B, while the other show less symmetry against $R$, and $G$. We also see that TB object and OG objects are most different from 0 in all three $(X-Y)$ combinations.

To determine the robustness of these \textit{cat}egories we examined the $(B-G)$ vs. $R$ color-magnitude in greater detail. Through the use of a k-means clustering algorithm and subsequent distortion and silhouette analyses \citep{sklearn}, we were able to sequentially test increasing numbers of clusters and determine the optimal number of clusters that best represented our object sample as has been previously done to select representatives from large samples \citep{Mayorga2019}. Examination of the distortion metric, ``the elbow curve'', revealed that by the inclusion of six or seven clusters the metric had dropped below 0.5, but our ``elbow'' was not sharp. With this method we were able to determine that the appropriate cluster number was six based on the \textit{cat}egories of $(B-G)$ vs. $R$. The distortion curve is similar for all three of the color-magnitude diagrams. 

\begin{figure}[!t]
    \epsscale{1.1}
    \plotone{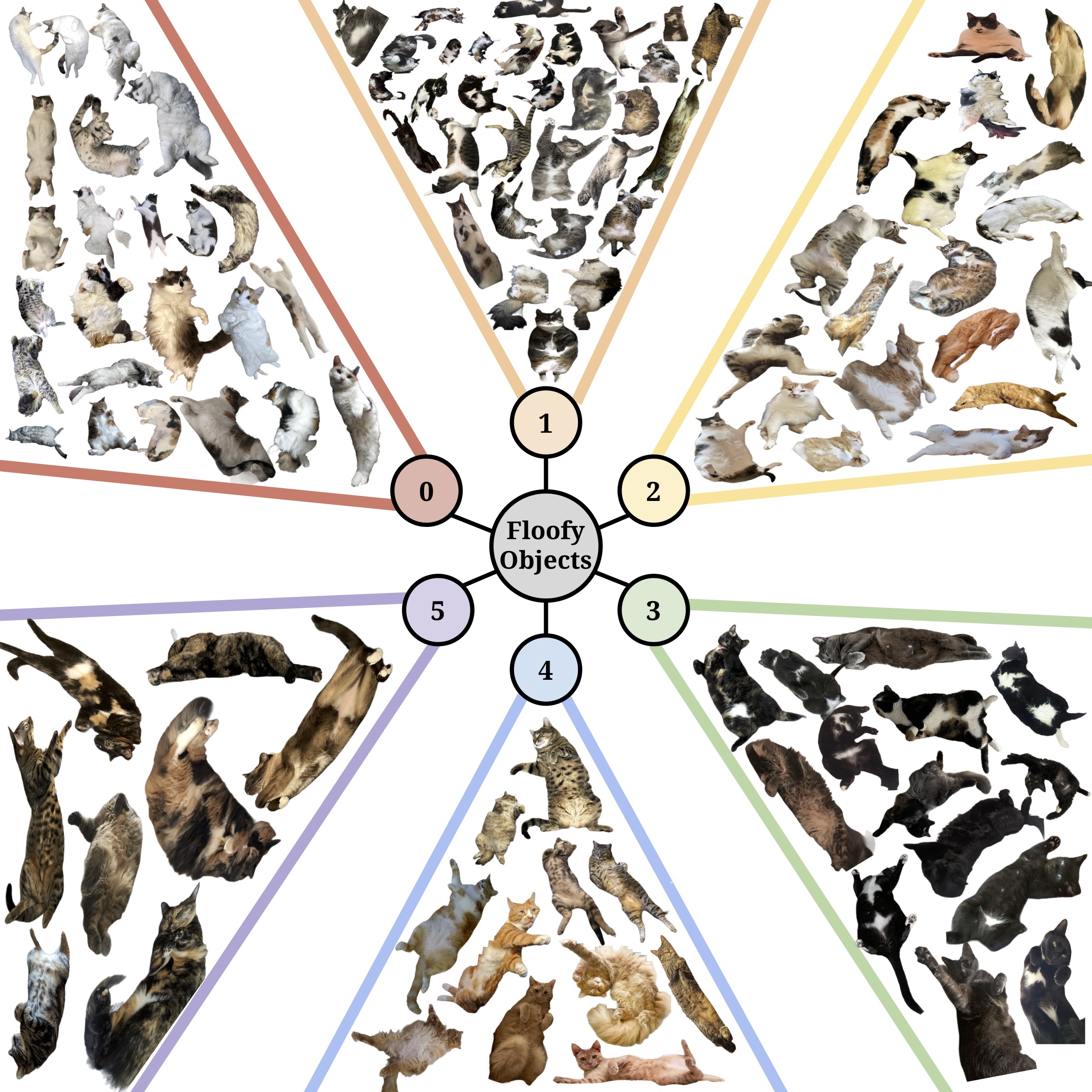}
    \caption{Here we show the 6 identified clusters and their members. Note that while clustering was done on the internally processed and white balanced frames, frames shown here are not white-balanced as we do not save these intermediate data processing steps as images. A higher resolution version of this figure can be made available upon request.}
    \label{fig:Clusters}
\end{figure}

Figure \ref{fig:Clusters} shows the cluster membership with non white-balanced longitude=0$^{\circ}$ frames\footnote{While clustering was done on the white-balanced frames, they are an intermediate data product that we do not save for all objects. For an example of the impact of white balancing, see Figure \ref{fig:cat_coordinates}.}. Visually we note that clusters 0, 1, and 2 are predominately bright at a longitude of 0, with differing amount of colors visible on their limbs. Cluster 0 contains many WT and CL cats with stronger limb brightness, while cluster 1 contains many BW sub-types with darker limbs. Cluster 2 appears to be an intermediate case with lighter colors on the limbs than cluster 1, but darker than cluster 0. Cluster 3 contains mostly BK sub-types with dark longitude=0$^{\circ}$ angles, continuing onto dark limbs. Cluster 4 and 5 contain predominantly TB and OG sub-types, with cluster 4 containing the lighter and brighter of the objects. In general, the first three clusters are the most rotationally variable, with the later three clusters more homogeneous. We conclude that the clustering algorithim used here is likely to be a powerful tool in \textit{cat}egorizing and understanding floofy objects, particularly for future observations with more optimal instrument set-ups.

\subsection{Shape and Rotation Rate}
Our floofy objects fall within the larger \textit{cat}egory of small bodies. Since objects were reported on phase angle, it is difficult to report on the rotation rate of the objects without temporal information. While the uniformity in albedo allows for shape and rotation rate to be readily recovered from small body light curves, planetary objects have an entanglement of both rotation and illumination variation effects. Resolved observations of floofy objects show that their spatial inhomogeneities are the main cause of the variation we see. Thus, floofy objects are more akin to planetary objects rather than small bodies. Therefore, we can apply mapping utilities as developed for planetary objects. This also means that determining the size of a floofy object is difficult without observing other objects for reference and comparison, as during transit and eclipse. This means it is difficult to determine an object's density and ensure it is not a super-puff \citep{Lopez2014, Masuda2014} or erroneously large such that it is a phantom inflated object \citep{Mayorga2018}.

\begin{figure}[t!]
    \centering
    \includegraphics[width=\linewidth]{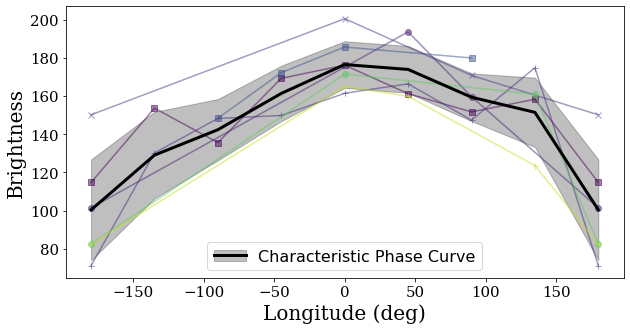}
    \includegraphics[width=0.9\linewidth, trim=-60 0 0 0, clip]{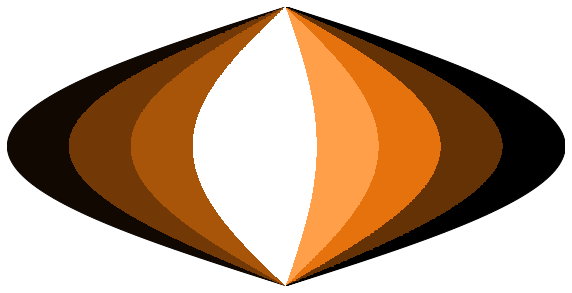}
    \caption{Characteristic rotation curve and inferred longitudinal map from our sample of sub\textit{cat}egory CL. }
    \label{fig:characteristic_CL}
\end{figure}

Our relatively large sample size of CL objects enables a deeper dive into the characteristics of this sub-sample and allows us to build a representative CL UFO. The upper panel of Figure \ref{fig:characteristic_CL} shows a characteristic rotation curve for CL objects, which we take to be the mean and standard deviation of all observations of this sub\textit{cat}egory at each longitude. The strong peak in brightness seen near $0^{\circ}$ longitude is clearly visible in the characteristic rotation curve along with the strong drop in brightness towards $180^{\circ}$. 

The lower panel of Figure \ref{fig:characteristic_CL} shows the longitudinal map that we infer from the characteristic rotation curve displayed on a sinusoidal map projection. To construct the map, we projected the longitudinally-dependent mean brightness from the upper panel onto a 9-slice longitudinal map using the Python \texttt{Basemap} code. The color of each slice corresponds to the mean brightness of CL objects at the given longitude, displayed using a custom linearly segmented colormap based on a crude prediction of how the objects would be perceived to the human eye. Similar rotational mapping techniques have been proposed as a means to detect oceans on rocky exoplanets \citep[e.g.][]{Cowan2009, Lustig-Yaeger2018}. 
Although the inferred brightness map assumes the unresolved objects are spherical in nature, we possess no evidence to the contrary and argue that radially symmetric gravitational forces will make such CL sub\textit{cat}egory objects spherical following oligarchic growth in the feeding zone.

\subsection{False \textbf{Paw}sitives}
The colors and rotational variability seen in the floofy objects identified in the study are not unique to this class of objects. For non-spatially resolved floofy objects, there is a potential for false \textit{paw}sitive confusion with WOOF Objects (Wagging tails On Objectively Friendly Objects). We have identified at least one potential false \textit{paw}sitive WOOF Object with a similar visual average color as the CL class of floofy objects, denoted DX01, for which we have obtained a half-rotation observation (data obtained through private communi\textit{cat}ion). Data reduction followed the same process as for the standard floofy objects.

\begin{figure}
    \centering
    \includegraphics[width = \linewidth]{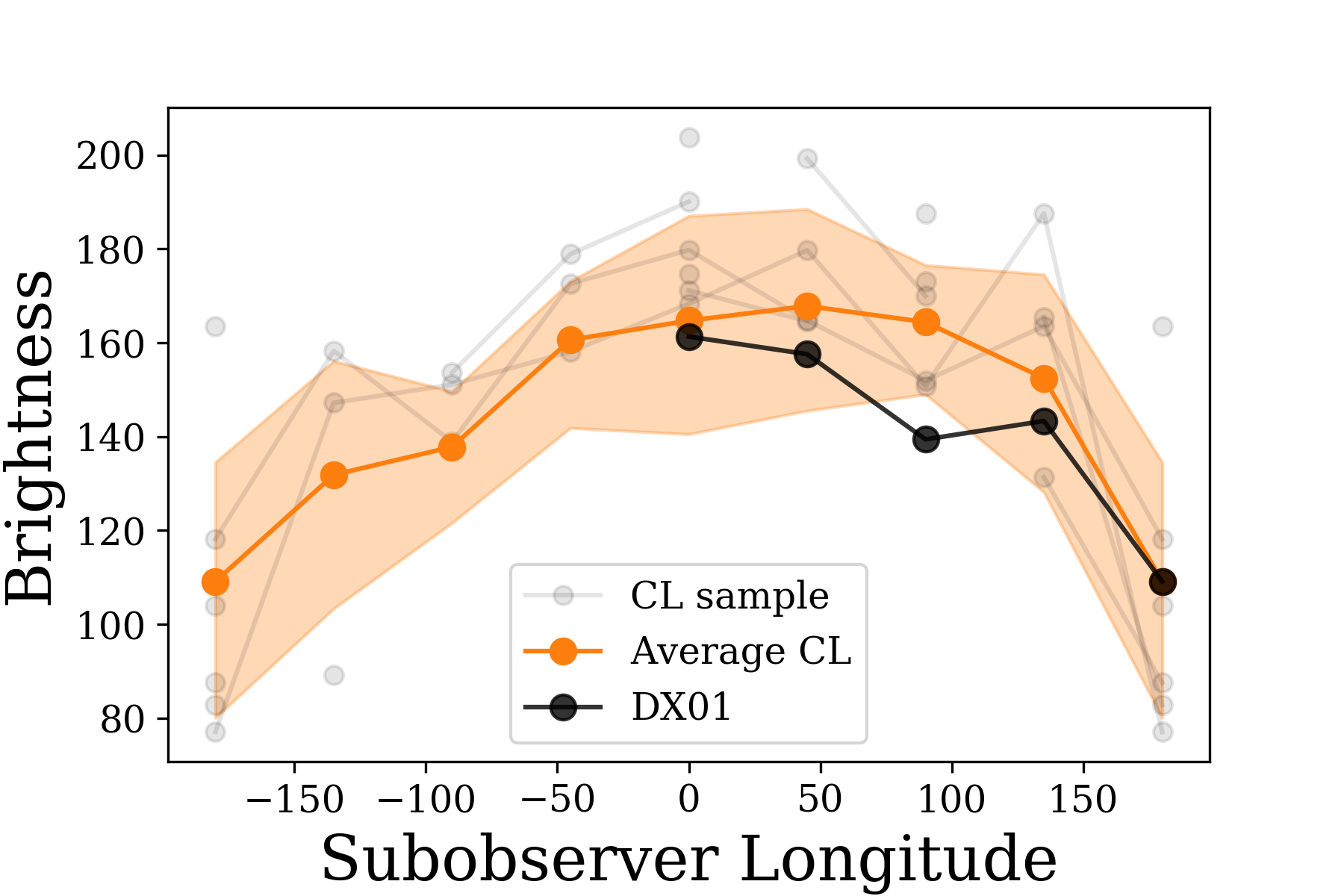}
    \caption{WOOF Object DX01 compared to the average CL rotation curve. DX01 has been normalized at a sub-observer longitude of 180 for comparison.}
    \label{fig:DX}
\end{figure}

In Figure \ref{fig:DX}, we compare the partial DX01 rotation curve to the average CL rotation curve. We find an initial vertical offset between WOOF Object DX01 and the average CL sub-type floofy object, which we attribute to variability in host object spot coverage. Visual assessment suggests that a sub-observer longitude of 180$^{\circ}$ is a similar color between DX01 and the average CL floofy object, so without further photometric monitoring we make the simplifying assumption to normalize DX01 to the mean CL sub-type rotation curve at a longitude of 180$^{\circ}$. We find that 4 out of 5 of the observations of DX01 fall within 1$\sigma$ of the average CL sub-type rotation curve, suggesting that for a uniform observational set up, the DX sub class is indistinguishable from the mean CL type object in unresolved observations. The potential for this false \textit{paw}sitive identifi\textit{cat}ion of CL sub-type floofy objects suggests that the CatEx high spatial resolution direct imaging mission is needed now more than ever, as it is simply unacceptable to misidentify WOOF Objects as Floofy Objects, particularly when they are known to have incredibly different orbital motions and temperaments.

\subsection{Additional Sources of Rotational Variability}
As hinted at in Figure \ref{fig:cat_coordinates}, floofy objects are more than the spherical cow approximation we have applied in this work. They appear to have a second sphere attached at their north pole, reminiscent of the shape of (486958) Arrokoth (provisional designation 2014 MU$_\mathrm{69}$), the most distant object observed in our Solar System. Unlike Arrokoth, this secondary sphere appears to rotate at a different rate than the rest of the floofy objects, sometimes changing direction unexpectedly. This additional source of variability in the color of floofy objects is a significant source of uncertainty and is not constrained in this work. 

Variability between observational epochs due to patchy cloud cover (and therefore changing reflectivity) has been predicted for tidally locked terrestrial worlds \citep{May2021} and sub-Neptunes \citep{Charnay2020}, though this is the first observational evidence for this epoch-to-epoch variability in rotationally inhomogeneous objects to-date. Further observations are required to understand how this effect is related to the secondary spherical component of floofy objects, and how the angle of this component contributes to the overall color and reflectivity of UFOs.

\section{Discussion} \label{sec:disc}


Generally speaking, for exoplanets, the best observational targets are those tidally locked, zero-obliquity Hot Jupiters due to their overwhelming benefits of size and contrast ratios compared to their smaller, cooler, planetary siblings. The large signal sizes for these planets mean they are among the only targets for which ground-based transit characterization can occur at high precision \citep[e.g.][]{May2018, May2020a, Sotzen2020}. However, in the context of understanding the relation between Solar System planets and the thousands of exoplanets, it is imperative we have nearly one-to-one observational comparisons. Because we currently lack the ability to watch the gaseous planets in our Solar System transit the Sun to derive comparative transmission observations, we can instead perform phase curve observations of exoplanets to compare to the reflected and emitted light phase curves observed of those planets closer to home.

Phase curves of exoplanets have primarily been observed with the Spitzer space telescope \citep[see ][and references therein]{Deming2020} to study planetary emission at infrared wavelengths, but data from planetary detection missions like TESS have also provided ever important optical phase curves directly measuring reflected light from these planets \citep[e.g.][]{Wong2020}. As we watch these planets orbit their host stars, their tidal locking, and therefore presumed lack of rotational obliquity, allows us to determine exactly which side of the planet we are seeing at any given point in time. Such determinations further enable us to directly measure differences between the day- and night sides of these worlds. Hot Jupiters typically show extreme variation between their day and night hemispheres due to the constant irradiation of their day sides, while the night side depends on the redistribution of heat in the atmosphere. This leads to a wealth of difference in cloud formation, and hence reflectivity, at optical wavelengths, as we view different sides of the exoplanet. 

In this day- and night side variance context, exoplanet observations suggest that in addition to longitudinal differences of rotationally variable objects, further study of floofy objects is warranted as a function of temporal phase. In addition, future work examining the effects of transit and eclipse of one optically variable floofy object sub\textit{cat}egory, such as CL or TW, with an optically nonvariable sub\textit{cat}egory, such as WT or BK, would reveal additional constraints into the behavior, morphology, and evolution of these floofy objects as a population. 
The upcoming launch of JWST may enable the methods pioneered here to extend down to smaller and fainter floofs, although we caution that multiple stacked observations may be necessary to characterize these objects in detail \citep{Lustig-Yaeger2019}. 

\subsection{\textbf{Cat}veats}

The majority of the observations were taken at 0$^\circ{}$ longitude. While this would appear to suggest that UFO prefer this orientation, our sample is biased by our specific request for floofy objects ``rolling around'' and that we had particular interest in floofy objects that exhibited brightness variations. This may have limited our sample, particularly in the BK, RD, and SM sub\textit{cat}egories, for which we received insufficient observations to produce a rotation curve. 

We do not have a sufficiently large sample to account for population completeness. Our small sample size also meant that we had to employ white-balancing techniques in order to account for variations in lighting as we also had insufficient illumination phase angle coverage. A future study could expand to study the phase variance in floofy objects as well as perhaps explore the topic in the lab where other conditions can be controlled. By bringing UFOs into the lab, their light curves and phase curves can be measured in a more controlled environment, however, it may be difficult to coerce them into the requisite viewing geometry. For example, an inherent assumption within our study is that our floofy objects are compositionally similar, though previous exoplanet-focused laboratory work has shown the potential pitfalls of such assumptions \citep{moran2020}.

\section{Conclusions} \label{sec:concl}
Our analysis of social media submitted observations of floofy objects has resulted in a sample of 328 objects, with 346 frames determined to be of sufficient quality for further analysis. Based on individual classifi\textit{cat}ion of resolved images, this resulted in the identifi\textit{cat}ion of 13 different sub\textit{cat}egories of floofy objects. 

We presented rotational curves of 32 different UFOs, which were well sampled in sub-observer longitude. There is a trend where at 0$^\circ{}$ longitude in UFOs is typically brighter and whiter than at any other sub-observer longitude. Research into morphological variations among more distant cousins of UFOs shows that this particular feature may be an environmental process that is sculpting the floof of our floofy objects with time \citep{Sanchez-Villagra2016}.

We also presented color-magnitude diagrams for our three filter combinations, $R$, $G$, and $B$. Based on a cluster analysis of $(B-G)$, $R$ we were able to determine that the observations, using solely 0$^\circ{}$ longitude observations were best represented by six sub\textit{cat}egories.

Under the assumption that our objects are spherical, and that color and brightness variations are due to spatial inhomogeneities, we were able to build a representative CL UFO and produce a brightness map for this sub\textit{cat}egory of floofy object.

\acknowledgements
We first acknowledge floofy objects in nearby orbits: L.C.M acknowledges Stellar; E.M.M acknowledges Button; S.E.M acknowledges Zorya and Set. J.L-Y acknowledges a WOOF Object, Moony. 

We further acknowledge Rex, Bo, SilmShady, JimmyJasper, Zoe, Murray, Mal, Megatron/Charles, Jimmy, Hobbes, Grasshopper, Admiral, FuzzyBritches, Nugget, Griffen, Schopenhauer, Ludwig, Dandy, Tadpole, Shiare, Bootsie, Amarielhin, Chuy, Aisling, Oliver, Emmy, Elliott, BFF, Velveteen, Trinity, Bruce, Geordi, Gracie, Chunky, Kepler, Onyx, Salem, Georgia, Jasper, Ellie, Poppy, Olive, Slippers, Sebastian, Teacup, Mowgli, Mitzi, Pinto, Freya, Luma, Merak, Mila, Olivia, Pompeya, Aries, Anya, Hulie, Lucy, Mina, Kirby, Remi, Boone, Buttercup, Max, Shadow, Sasha, Eleanor, Teryx, Wadsworth, Doug, Kaylee, Schnurri, Cricket, Quiche, Spencer, Indy, ZsaZsa, Slyvester, Fred, Challa, Finn, Nimbus, Sara, Millie, Olive, Willow, Anya, Kilo, Whisper, Vern, Oliver, Allie, Eris, ChumChum, FelixFjord, Peanut, Eleanor, Jake (not to be confused with the \textit{definitely not a cat} author Jake), Petals, Ms.Moon, Sleeves, Smog, Cotton, Athena, Nala, Lovebug, Bigsby, Marcel, Tanya, Albus, HoneyJumbles, Hansel, Henri, BathroomCat, Addie, Prince, Milo, Ziggy, Happy, PB, Henri, Layla, Prince, Damian, Milly, Dixie, Mishti, Lily, Taters, Clem, Marley, Gildor, Renato, Nemesis, Myrtle, Boylar, Ivy, Nico, Lester, MamaCat, Donut, Fiyero, Astrid, Andy, Bob, Diesel, Luna, PercyJulian, Brad, Mugzle, Lulu, MissOlivia, Lucy, Merlin, Sawyer, Madmartigan, PinotNoir, Louis, Charlotte, Perry, Kevin, Cheeto, Scallop, Taz, Misha, Sassy, Tio, Saturn, Simon, Mr.Cat, Isabelle, Joowol, Neelix, Kiki, Buffy, Stella, Sebastian, Eric, Pantalaimon, Mia, Nova, Wizard, Nova, Puff, Bo, Pierogi, Lucky, Whiskers, Freddie, Fenyman, Simcoe and 133 other unnamed floofy objects for their willingness to show their bellies for science.

\software{NumPy \citep{NumPy, numpynew}, Matplotlib (including basemap) \citep{matplotlib}, Pandas \citep{Pandas}, SciPy \citep{SciPy}, skimage \citep{skimage}}

\bibliographystyle{aasjournal}
\bibliography{refs}

\begin{thebibliography}{}
\expandafter\ifx\csname natexlab\endcsname\relax\def\natexlab#1{#1}\fi
\providecommand{\url}[1]{\href{#1}{#1}}
\providecommand{\dodoi}[1]{doi:~\href{http://doi.org/#1}{\nolinkurl{#1}}}
\providecommand{\doeprint}[1]{\href{http://ascl.net/#1}{\nolinkurl{http://ascl.net/#1}}}
\providecommand{\doarXiv}[1]{\href{https://arxiv.org/abs/#1}{\nolinkurl{https://arxiv.org/abs/#1}}}

\bibitem[{{Batalha} {et~al.}(2018){Batalha}, {Smith}, {Lewis}, {Marley},
  {Fortney}, \& {Macintosh}}]{Batalha2018}
{Batalha}, N.~E., {Smith}, A. J.~R.~W., {Lewis}, N.~K., {et~al.} 2018, \aj,
  156, 158, \dodoi{10.3847/1538-3881/aad59d}

\bibitem[{{Buratti} {et~al.}(2017){Buratti}, {Hofgartner}, {Hicks}, {Weaver},
  {Stern}, {Momary}, {Mosher}, {Beyer}, {Verbiscer}, {Zangari}, {Young},
  {Lisse}, {Singer}, {Cheng}, {Grundy}, {Ennico}, \& {Olkin}}]{buratti2017}
{Buratti}, B.~J., {Hofgartner}, J.~D., {Hicks}, M.~D., {et~al.} 2017, \icarus,
  287, 207, \dodoi{10.1016/j.icarus.2016.11.012}

\bibitem[{Burggraaff {et~al.}(2019)Burggraaff, Schmidt, Zamorano, Pauly,
  Pascual, Tapia, Spyrakos, \& Snik}]{Burggraaff:19}
Burggraaff, O., Schmidt, N., Zamorano, J., {et~al.} 2019, Opt. Express, 27,
  19075, \dodoi{10.1364/OE.27.019075}

\bibitem[{Cahoy {et~al.}(2010)Cahoy, Marley, \& Fortney}]{Cahoy2010}
Cahoy, K.~L., Marley, M.~S., \& Fortney, J.~J. 2010, Astrophys. J., 724, 189,
  \dodoi{10.1088/0004-637X/724/1/189}

\bibitem[{Charnay {et~al.}(2020)Charnay, Blain, B\'{e}zard, \& {et
  al.}}]{Charnay2020}
Charnay, B., Blain, D., B\'{e}zard, B., \& {et al.} 2020, arXiv
  e-prints:2011.11553

\bibitem[{{Cowan} {et~al.}(2009){Cowan}, {Agol}, {Meadows}, {Robinson},
  {Livengood}, {Deming}, {Lisse}, {A'Hearn}, {Wellnitz}, {Seager},
  {Charbonneau}, \& {EPOXI Team}}]{Cowan2009}
{Cowan}, N.~B., {Agol}, E., {Meadows}, V.~S., {et~al.} 2009, \apj, 700, 915,
  \dodoi{10.1088/0004-637X/700/2/915}

\bibitem[{{Deming} \& {Knutson}(2020)}]{Deming2020}
{Deming}, D., \& {Knutson}, H.~A. 2020, Nature Astronomy, 4, 453,
  \dodoi{10.1038/s41550-020-1100-9}

\bibitem[{Eizirik {et~al.}(2010)Eizirik, David, Buckley-Beason, Roelke,
  Schäffer, Hannah, Narfström, O'Brien, \& Menotti-Raymond}]{CatGenes1}
Eizirik, E., David, V.~A., Buckley-Beason, V., {et~al.} 2010, Genetics, 184,
  267, \dodoi{10.1534/genetics.109.109629}

\bibitem[{{Grundy} {et~al.}(2016){Grundy}, {Binzel}, {Buratti}, {Cook},
  {Cruikshank}, {Dalle Ore}, {Earle}, {Ennico}, {Howett}, {Lunsford}, {Olkin},
  {Parker}, {Philippe}, {Protopapa}, {Quirico}, {Reuter}, {Schmitt}, {Singer},
  {Verbiscer}, {Beyer}, {Buie}, {Cheng}, {Jennings}, {Linscott}, {Parker},
  {Schenk}, {Spencer}, {Stansberry}, {Stern}, {Throop}, {Tsang}, {Weaver},
  {Weigle}, \& {Young}}]{grundy2016}
{Grundy}, W.~M., {Binzel}, R.~P., {Buratti}, B.~J., {et~al.} 2016, Science,
  351, aad9189, \dodoi{10.1126/science.aad9189}

\bibitem[{{Hapke}(2012)}]{Hapke2012}
{Hapke}, B. 2012, \icarus, 221, 1079, \dodoi{10.1016/j.icarus.2012.10.022}

\bibitem[{Harris {et~al.}(2020)Harris, Millman, van~der Walt, Gommers,
  Virtanen, Cournapeau, Wieser, Taylor, Berg, Smith, Kern, Picus, Hoyer, van
  Kerkwijk, Brett, Haldane, del R{'{\i}}o, Wiebe, Peterson,
  G{'{e}}rard-Marchant, Sheppard, Reddy, Weckesser, Abbasi, Gohlke, \&
  Oliphant}]{numpynew}
Harris, C.~R., Millman, K.~J., van~der Walt, S.~J., {et~al.} 2020, Nature, 585,
  357, \dodoi{10.1038/s41586-020-2649-2}

\bibitem[{Hegde \& Kaltenegger(2013)}]{Hegde2013}
Hegde, S., \& Kaltenegger, L. 2013, Astrobiology, 13, 47,
  \dodoi{10.1089/ast.2012.0849}

\bibitem[{{Henyey} \& {Greenstein}(1941)}]{Henyey1941}
{Henyey}, L.~G., \& {Greenstein}, J.~L. 1941, \apj, 93, 70,
  \dodoi{10.1086/144246}

\bibitem[{Hunter(2007)}]{matplotlib}
Hunter, J.~D. 2007, Comput. Sci. Eng., 9, 90, \dodoi{10.1109/MCSE.2007.55}

\bibitem[{Jones {et~al.}(2001)Jones, Oliphant, Peterson, \& Others}]{SciPy}
Jones, E., Oliphant, T., Peterson, P., \& Others. 2001, {{\{}SciPy{\}}: Open
  source scientific tools for {\{}Python{\}}}.
\newblock \url{http://www.scipy.org/}

\bibitem[{Kaelin {et~al.}(2012)Kaelin, Xu, Hong, David, McGowan,
  Schmidt-K{\"u}ntzel, Roelke, Pino, Pontius, Cooper, Manuel, Swanson, Marker,
  Harper, van Dyk, Yue, Mullikin, Warren, Eizirik, Kos,
  O{\textquoteright}Brien, Barsh, \& Menotti-Raymond}]{CatGenes2}
Kaelin, C.~B., Xu, X., Hong, L.~Z., {et~al.} 2012, Science, 337, 1536,
  \dodoi{10.1126/science.1220893}

\bibitem[{{Lacerda} \& {Jewitt}(2007)}]{Lacerda2007}
{Lacerda}, P., \& {Jewitt}, D.~C. 2007, \aj, 133, 1393, \dodoi{10.1086/511772}

\bibitem[{{Lopez} \& {Fortney}(2014)}]{Lopez2014}
{Lopez}, E.~D., \& {Fortney}, J.~J. 2014, \apj, 792, 1,
  \dodoi{10.1088/0004-637X/792/1/1}

\bibitem[{{Lustig-Yaeger} {et~al.}(2019){Lustig-Yaeger}, {Meadows}, \&
  {Lincowski}}]{Lustig-Yaeger2019}
{Lustig-Yaeger}, J., {Meadows}, V.~S., \& {Lincowski}, A.~P. 2019, \aj, 158,
  27, \dodoi{10.3847/1538-3881/ab21e0}

\bibitem[{{Lustig-Yaeger} {et~al.}(2018){Lustig-Yaeger}, {Meadows}, {Tovar
  Mendoza}, {Schwieterman}, {Fujii}, {Luger}, \&
  {Robinson}}]{Lustig-Yaeger2018}
{Lustig-Yaeger}, J., {Meadows}, V.~S., {Tovar Mendoza}, G., {et~al.} 2018, \aj,
  156, 301, \dodoi{10.3847/1538-3881/aaed3a}

\bibitem[{{Masuda}(2014)}]{Masuda2014}
{Masuda}, K. 2014, \apj, 783, 53, \dodoi{10.1088/0004-637X/783/1/53}

\bibitem[{{May} {et~al.}(2020){May}, {Gardner}, {Rauscher}, \&
  {Monnier}}]{May2020a}
{May}, E.~M., {Gardner}, T., {Rauscher}, E., \& {Monnier}, J.~D. 2020, \aj,
  159, 7, \dodoi{10.3847/1538-3881/ab5361}

\bibitem[{{May} \& {Rauscher}(2016)}]{May2016}
{May}, E.~M., \& {Rauscher}, E. 2016, \apj, 826, 225,
  \dodoi{10.3847/0004-637X/826/2/225}

\bibitem[{{May} \& {Rauscher}(2020)}]{May2020b}
---. 2020, \apj, 893, 161, \dodoi{10.3847/1538-4357/ab838b}

\bibitem[{{May} \& {Stevenson}(2020)}]{May2020c}
{May}, E.~M., \& {Stevenson}, K.~B. 2020, \aj, 160, 140,
  \dodoi{10.3847/1538-3881/aba833}

\bibitem[{{May} {et~al.}(2021){May}, {Taylor}, {Komacek}, {Line}, \&
  {Parmentier}}]{May2021}
{May}, E.~M., {Taylor}, J., {Komacek}, T.~D., {Line}, M.~R., \& {Parmentier},
  V. 2021, arXiv e-prints, arXiv:2103.09313.
\newblock \doarXiv{2103.09313}

\bibitem[{{May} {et~al.}(2018){May}, {Zhao}, {Haidar}, {Rauscher}, \&
  {Monnier}}]{May2018}
{May}, E.~M., {Zhao}, M., {Haidar}, M., {Rauscher}, E., \& {Monnier}, J.~D.
  2018, \aj, 156, 122, \dodoi{10.3847/1538-3881/aad4a8}

\bibitem[{{Mayorga} {et~al.}(2019){Mayorga}, {Batalha}, {Lewis}, \&
  {Marley}}]{Mayorga2019}
{Mayorga}, L.~C., {Batalha}, N.~E., {Lewis}, N.~K., \& {Marley}, M.~S. 2019,
  \aj, 158, 66, \dodoi{10.3847/1538-3881/ab29fa}

\bibitem[{{Mayorga} {et~al.}(2020){Mayorga}, {Charbonneau}, \&
  {Thorngren}}]{Mayorga2020}
{Mayorga}, L.~C., {Charbonneau}, D., \& {Thorngren}, D.~P. 2020, \aj, 160, 238,
  \dodoi{10.3847/1538-3881/abb8df}

\bibitem[{Mayorga {et~al.}(2016)Mayorga, Jackiewicz, Rages, West, Knowles,
  Lewis, \& Marley}]{Mayorga2016}
Mayorga, L.~C., Jackiewicz, J., Rages, K., {et~al.} 2016, Astron. J., 152, 209,
  \dodoi{10.3847/0004-6256/152/6/209}

\bibitem[{Mayorga \& Thorngren(2018)}]{Mayorga2018}
Mayorga, L.~C., \& Thorngren, D.~P. 2018, Res. Notes AAS, 2, 40,
  \dodoi{10.3847/2515-5172/aac728}

\bibitem[{McKinney(2010)}]{Pandas}
McKinney, W. 2010, in Proc. 9th Python Sci. Conf., ed. S.~van~der Walt \&
  J.~Millman, 56--61, \dodoi{10.25080/Majora-92bf1922-00a}

\bibitem[{McReynolds \& Blythe(2005)}]{McREYNOLDS2005}
McReynolds, T., \& Blythe, D. 2005, in Adv. Graph. Program. Using OpenGL
  (Elsevier), 35--56, \dodoi{10.1016/B978-155860659-3.50005-6}

\bibitem[{{Moran} {et~al.}(2018){Moran}, {H{\"o}rst}, {Batalha}, {Lewis}, \&
  {Wakeford}}]{moran2018}
{Moran}, S.~E., {H{\"o}rst}, S.~M., {Batalha}, N.~E., {Lewis}, N.~K., \&
  {Wakeford}, H.~R. 2018, \aj, 156, 252, \dodoi{10.3847/1538-3881/aae83a}

\bibitem[{{Moran} {et~al.}(2020){Moran}, {H{\"o}rst}, {Vuitton}, {He}, {Lewis},
  {Flandinet}, {Moses}, {North}, {Orthous-Daunay}, {Sebree}, {Wolters},
  {Kempton}, {Marley}, {Morley}, \& {Valenti}}]{moran2020}
{Moran}, S.~E., {H{\"o}rst}, S.~M., {Vuitton}, V., {et~al.} 2020, The Planetary
  Science Journal, 1, 17, \dodoi{10.3847/PSJ/ab8eae}

\bibitem[{Pedregosa {et~al.}(2012)Pedregosa, Varoquaux, Gramfort, Michel,
  Thirion, Grisel, Blondel, M{\"{u}}ller, Nothman, Louppe, Prettenhofer, Weiss,
  Dubourg, Vanderplas, Passos, Cournapeau, Brucher, Perrot, \&
  Duchesnay}]{sklearn}
Pedregosa, F., Varoquaux, G., Gramfort, A., {et~al.} 2012, J. Mach. Learn.
  Res., 12, 2825, \dodoi{10.1007/s13398-014-0173-7.2}

\bibitem[{{Rasio} {et~al.}(1996){Rasio}, {Tout}, {Lubow}, \&
  {Livio}}]{Rasio1996}
{Rasio}, F.~A., {Tout}, C.~A., {Lubow}, S.~H., \& {Livio}, M. 1996, \apj, 470,
  1187, \dodoi{10.1086/177941}

\bibitem[{S\`anchez-Villagra {et~al.}(2016)S\`anchez-Villagra, Geiger, \&
  Schneider}]{Sanchez-Villagra2016}
S\`anchez-Villagra, M., Geiger, M., \& Schneider, R. 2016, Royal Society Open
  Science, 3, 160107, \dodoi{10.1098/rsos.160107}

\bibitem[{{Sotzen} \& {et al.}(Submitted, 2021)}]{Sotzen2021}
{Sotzen}, K.~S., \& {et al.} Submitted, 2021, ApJ

\bibitem[{{Sotzen} {et~al.}(2020){Sotzen}, {Stevenson}, {Sing}, {Kilpatrick},
  {Wakeford}, {Filippazzo}, {Lewis}, {H{\"o}rst}, {L{\'o}pez-Morales}, {Henry},
  {Buchhave}, {Ehrenreich}, {Fraine}, {Garc{\'\i}a Mu{\~n}oz}, {Jayaraman},
  {Lavvas}, {Lecavelier des Etangs}, {Marley}, {Nikolov}, {Rathcke}, \&
  {Sanz-Forcada}}]{Sotzen2020}
{Sotzen}, K.~S., {Stevenson}, K.~B., {Sing}, D.~K., {et~al.} 2020, \aj, 159, 5,
  \dodoi{10.3847/1538-3881/ab5442}

\bibitem[{{Travis E}(2006)}]{NumPy}
{Travis E}, O. 2006, {A guide to NumPy},  USA: Trelgol Publishing

\bibitem[{Union(2007)}]{international2007studio}
Union, I.~T. 2007, Studio Encoding Parameters of Digital Television for
  Standard 4:3 and Wide-screen 16:9 Aspect Ratios: Recommendation ITU-R
  BT.601-6 : (Question ITU-R 1/6). (International Telecommunication Union).
\newblock \url{https://books.google.com/books?id=NpAwPwAACAAJ}

\bibitem[{van~der Walt {et~al.}(2014)van~der Walt, Sch{\"{o}}nberger,
  Nunez-Iglesias, Boulogne, Warner, Yager, Gouillart, \& Yu}]{skimage}
van~der Walt, S., Sch{\"{o}}nberger, J.~L., Nunez-Iglesias, J., {et~al.} 2014,
  PeerJ, 2, e453, \dodoi{10.7717/peerj.453}

\bibitem[{{Wong} {et~al.}(2020){Wong}, {Shporer}, {Daylan}, {Benneke},
  {Fetherolf}, {Kane}, {Ricker}, {Vanderspek}, {Latham}, {Winn}, {Jenkins},
  {Boyd}, {Glidden}, {Goeke}, {Sha}, {Ting}, \& {Yahalomi}}]{Wong2020}
{Wong}, I., {Shporer}, A., {Daylan}, T., {et~al.} 2020, \aj, 160, 155,
  \dodoi{10.3847/1538-3881/ababad}

\end{thebibliography}

\end{document}